\documentclass[prd, aps, reprint, amsfonts, amssymb, amsmath, preprintnumbers, showpacs, nofootinbib, superscriptaddress]{revtex4-2}
\usepackage{graphicx, multirow,soul}
\usepackage[citecolor=blue]{hyperref}
\usepackage[all]{hypcap}
\usepackage{url,ulem}
\usepackage{slashed}
\usepackage{lipsum}
\usepackage[table]{xcolor}

\begin{document}

\title{Classification of Abelian domain walls}

\author{Yongcheng Wu}
\email{ycwu@njnu.edu.cn}
\affiliation{Department of Physics and Institute of Theoretical Physics, Nanjing Normal University, Nanjing, 210023, China}
\author{Ke-Pan Xie}
\email{kpxie666@163.com}
\affiliation{Department of Physics and Astronomy, University of Nebraska, Lincoln, NE 68588, USA}
\author{Ye-Ling Zhou}
\email{zhouyeling@ucas.ac.cn (corresponding author)}
\affiliation{School of Fundamental Physics and Mathematical Sciences, Hangzhou Institute for Advanced Study, UCAS, Hangzhou, 310024 China}
\affiliation{International Centre for Theoretical Physics Asia-Pacific, Beijing/Hangzhou, China}

\date{\today}

\begin{abstract}
We discuss domain walls from spontaneous breaking of Abelian discrete symmetries $Z_N$. A series of different domain wall structures are predicted, depending on the symmetry and charge assignments of scalars leading to the spontaneous symmetry breaking (SSB). A widely-existing type of domain walls are those separating degenerate vacua which are adjacent in the field space. We denote these walls as adjacency walls. In the case that $Z_N$ terms are small compared with the $U(1)$ terms, the SSB of $U(1)$ generates strings first and then adjacency walls bounded by strings are generated after the SSB of $Z_N$. For symmetries larger than $Z_3$, non-adjacent vacua exist, we regard walls separating them  as non-adjacency walls. These walls are unstable if $U(1)$ is a good approximation. If the discrete symmetry is broken via multiple steps, we arrive at a complex structure that one kind of walls wrapped by another type. On the other hand, if the symmetry is broken in different directions independently, walls generated from the different breaking chains are blind to each other.

\end{abstract}
\preprint{}
\maketitle

\section{Introduction}

Discrete symmetries have been widely discussed in particle physics. In the Standard Model (SM), the CP symmetry, which is an approximate $Z_2$ symmetry, is one of the most famous examples. Discrete symmetries are also frequently applied in new physics models to restrict interactions at high energy scales. Typical examples include, e.g., the matter parity in left-right symmetric and $SO(10)$ grand unified models \cite{Kibble:1982dd,Chang:1983fu}, the R-parity in supersymmetric models~\cite{Ibanez:1991pr}, the $Z_N$ symmetry in axion-like particle models~\cite{Sikivie:1982qv}, the Abelian and/or non-Abelian discrete symmetries in flavour models~\cite{King:2017guk,Xing:2019vks}, and modular symmetries arising from formal theories \cite{Ferrara:1989bc,Ferrara:1989qb}.

In many cases, the discrete symmetries are spontaneously broken by the nontrivial vacuum structure of the system. Namely, even though the scalar potential itself is invariant under the discrete symmetry, if it has a set of degenerate vacua that are not invariant under the symmetry, then the spontaneous symmetry breaking (SSB) exists when the system stays in one specific vacuum. Due to the thermal corrections, discrete symmetries are generally restored in the very early Universe. During the cooling of the Universe along the Hubble expansion, SSB exists. As there is no preference in the degenerate vacua, spatial regions not connected by causality are free to nucleate to any vacua, resulting in a multi-bubble Universe with different cells staying in different vacua. The boundaries of the cells are two dimensional topological solitons called domain walls \cite{Kibble:1976sj}.

Domain walls are in general regarded as a problem in cosmology: their energy density decreases more slowly than the radiation and cold matter energy densities, thus they may dominate the Universe at late time, leading to an accelerated expansion era which is ruled out by observations \cite{Zeldovich:1974uw}. To solve this problem, one may either push the SSB of discrete symmetries earlier than the inflation or introduce explicit-breaking terms in the potential  \cite{Larsson:1996sp,Gelmini:1988sf,Vilenkin:1981zs} (see, e.g., \cite{Stojkovic:2005zh, King:2018fke} for other possible ways). These terms generate biases between the vacua. They later become significant as the Universe cools down to a temperature sufficiently lower than the scale of SSB. Then vacua with higher energy become unstable and domain walls collapse, avoiding the cosmological problem. The collapsing domain walls lead to the production of gravitational waves (GWs), which form a stochastic background today, providing us a way to test discrete symmetries of particle physics.

In the literature, most studies on domain wall evolution and GW productions focus on the classical $Z_2$ domain walls as an illustrative example. See Ref.~\cite{Saikawa:2017hiv} for a recent review, and lattice simulations are performed in Refs.~\cite{Hiramatsu:2010yz,Hiramatsu:2013qaa}. Numerical simulations on string-wall networks in $Z_N$-invariant axion models are performed in Refs.~\cite{Hiramatsu:2012sc,Kawasaki:2014sqa}, and that for a generic set of potentials is recently discussed in \cite{Krajewski:2021jje}. Phenomenological applications of domain wall-induced GWs as a way to test new physics have caught attentions, e.g., those applied to spontaneous R-parity symmetry breaking using GWs \cite{Dine:2010eb}, spontaneous CP violation at scalar-extended electroweak symmetry breaking \cite{Chen:2020wvu,Chen:2020soj}, $Z_3$-invariant singlet-scalar-extended SM~\cite{Zhou:2020ojf}, domain-wall-seeding electroweak phase transition~\cite{Blasi:2022woz}, and Dirac leptogenesis in $Z_2$ symmetry \cite{Barman:2022yos}. GW detections could therefore provide a characteristic signature for a large range of spontaneous discrete symmetry breaking scales that are hard to be tested in other experiments. Ref. \cite{Gelmini:2020bqg} points out that GW signal induced by domain walls provides a potential way to test the origin of lepton flavour mixing. A large range of discrete flavour symmetry scale, from 1~TeV to $10^{14}$~GeV, can be potentially touched by the next-generation GW interferometers, depending on an adequately-chosen bias parameter. Axion-like particles as a dark matter candidate can be detected with mass from $10^{-16}$ to $10^6$~eV if they are produced at temperatures below 100 eV \cite{Gelmini:2021yzu}.

In the last two years, Pulsar Timing Array (PTA) observatories, including NANOGrav \cite{Brazier:2019mmu}, European PTA \cite{Desvignes:2016yex}, Parkes PTA \cite{Kerr:2020qdo}, have reported evidences for a common-spectrum process in the search of GW background \cite{NANOGrav:2020bcs, Goncharov:2021oub, Chen:2021rqp}, and their individual data was reinforced in the analysis of International PTA \cite{Antoniadis:2022pcn}.
These signals have been interpreted as a hint of GWs from cosmic domain walls in either $Z_2$ \cite{Arzoumanian:2020vkk} or in the axion-like-particle models \cite{Bian:2020urb, Sakharov:2021dim}, or in a model-independent analysis \cite{Ferreira:2022zzo}.

Recently we explored domain wall properties and the consequent GW signatures from discrete symmetries beyond $Z_2$ \cite{Wu:2022stu}. A thorough study on $Z_3$ domain walls has been taken for illustration, where semi-analytical results for the tension and thickness of domain walls are derived. We pointed out that, as multiple degenerate vacua exist in the theory, an explicit breaking term leads to multiple biases between the vacua. Due to these biases, domain walls separating different vacua collapse at different time in the early Universe, and the process of domain wall collapsing is more complicated than the simplest $Z_2$ case. As a consequence, GW spectrum from these walls is different from those in the $Z_2$ case.

In this paper, we continue on the exploration of domain wall properties from Abelian discrete symmetries beyond $Z_2$, particularly focusing on classification of $Z_N$ domain walls. Compared with earlier discussions on this topic \cite{Vilenkin:2000jqa},  We will carry out a broader study on these topological defects from $Z_N$ symmetries for $N=3$, 4, 5, 6. A clarification of domain wall structures for general $Z_N$ will be made. The rest of this paper is organised as follows. We list scalar potentials for different $Z_N$ symmetries in section~\ref{sec:potential}. These potential are critical for domain wall formation. In section~\ref{sec:small_Z_N}, we discuss properties of domain walls from $Z_N$ with $N \leqslant 4$. Section~\ref{sec:large_Z_N} contributes to domain walls formed in larger $Z_N$ symmetries such as $Z_5$ and $Z_6$. General remarks for domain wall categories formed in $Z_N$ symmetries will be discussed in the end of the same section. We summarise and conclude in section~\ref{sec:summary}. This paper focus on the classification of $Z_N$ domain wall structures and thus no explicit breaking (e.g. energy bias terms) will be discussed. We also assume the CP symmetry is a good symmetry and do not consider any effect referring to the SSB of CP in the whole paper.

\section{$Z_N$-invariant potentials}\label{sec:potential}

The scalar potential can be determined by either one single complex scalar or multiple scalars. We discuss these two possibilities in the following two subsections.

\subsection{Potential with a single complex scalar}

We begin with the classical $Z_2$ case. The simplest way to describe the SSB of a $Z_2$-invariant theory is considering the renormalisable potential of a real scalar $h$ ($h \to -h$ under $Z_2$ transformation) as
\begin{eqnarray} \label{eq:potential_Z2}
V_{Z_2} = - \frac{\mu^2}{2} h^2 + \frac{\lambda}{4} h^4\,,
\end{eqnarray}
where both $\mu$ and $\lambda$ are real and positive. The minima of $V(\phi)$ appears at $\frac{h}{\sqrt{2}} = v_0$, $v_1$ with $v_{0,1} = \pm \sqrt{\mu^2/(2\lambda)}$.

For $Z_N$ with $N \geqslant 3$, a real scalar is not enough.
Given a complex scalar $\phi = \frac{1}{\sqrt{2}} (h+ ia)$, the symmetry requires the whole theory is invariant under the transformation
\begin{eqnarray}
T: \,\phi \to e^{i \frac{2\pi }{N} q_\phi} \phi\,.
\end{eqnarray}
where $q_\phi$ is the charge of $\phi$ in $Z_N$. Note that if there is a non-trivial greatest common divisor (gcd) between $q_\phi$ and $N$, i.e., ${\rm gcd}\{ q_\phi, N \} = n >1$, the essential symmetry where $\phi$ is evolving is $Z_{N/n}$. Thus, it is enough to consider just the case ${\rm gcd}\{ q_\phi, N \} = 1$, i.e., $q_\phi$ coprime with $N$. Without loss of generality, we can fix the charge $q_\phi =1$. $Z_N$-invariant operators of $\phi$ must be $\phi^* \phi$, $\phi^N$, $\phi^{* N}$ or their combinations. Imposing the CP symmetry,
\begin{eqnarray}
S: \,\phi \to \phi^*\,,
\end{eqnarray}
$\phi^N$ and $\phi^{* N}$ are enforced to appear as combinations of $\phi^N+\phi^{* N}$. The transformations $T$ and $S$ satisfy $T^N = S^2 = (TS)^2 =1$. While $T$ results in a rotation of $2\pi/N$ on the complex plane of the field $\phi$, $S$ is a reflection between the positive and negative imaginary parts. They generate the dihedral group $D_N$ which is isomorphic to $Z_N \rtimes Z_2^{CP}$ and also denoted as $\Delta(2N)$. Here, the parity symmetry $Z_2^{CP}$ represents the CP symmetry imposed in the theory. For the group theories of $D_N$, see e.g., Ref. \cite{Ishimori:2010au}.  We emphasise that $D_N$ is a natural consequence of the Abelian discrete symmetry $Z_N$ and the CP symmetry.

In general, the $Z_N$- and CP-invariant potential for a complex scalar $\phi$ must take the form
\begin{eqnarray} \label{eq:potential_general}
{\cal V}_{Z_N} = f(\phi^* \phi, \phi^N +\phi^{* N}) \,.
\end{eqnarray}
Here and below, we do not consider spontaneous CP violation, which can be achieved with appropriate coefficients arranged. Thus, one of the vacua can be chosen to be real, and we can further keep it positive with the help of a redefinition of the field $\phi \to -\phi$. We denote this real and positive vacuum as $v_0$. It is obtained by solving the equation
\begin{eqnarray}
\partial_x f(x,y) + N v_0^{N-2} \partial_y f(x,y)\Big|_{\{x, y\}=\{v_0^2,2 v_0^N\}} =0\,.
\end{eqnarray}
The rest vacua are obtained by the $T$ transformation of $Z_N$, $v_k = T^k v_0$.  Then, we arrive at $N$ degenerate vacua, i.e., $\langle\phi\rangle=v_k$ and
\begin{eqnarray}
v_k = v_0 \, e^{i \frac{2\pi}{N} k},
\end{eqnarray}
for $k=0,1,..., N-1$. The $S$ transformation does not give us any additional vacua. In the complex plane $(h, a)$, the potential in Eq.~\eqref{eq:potential_general} takes a general feature of {\it the bottom of a classical coca-cola bottle}. That is the local maximal point at $\phi =0$ in the central surrounded by $N$ minima separated by an angle $2\pi/N$. We will show a few examples in Fig.~\ref{fig:potential_ZN}, and the details are given below.

\begin{figure*}
\centering
\includegraphics[width=.6\textwidth]{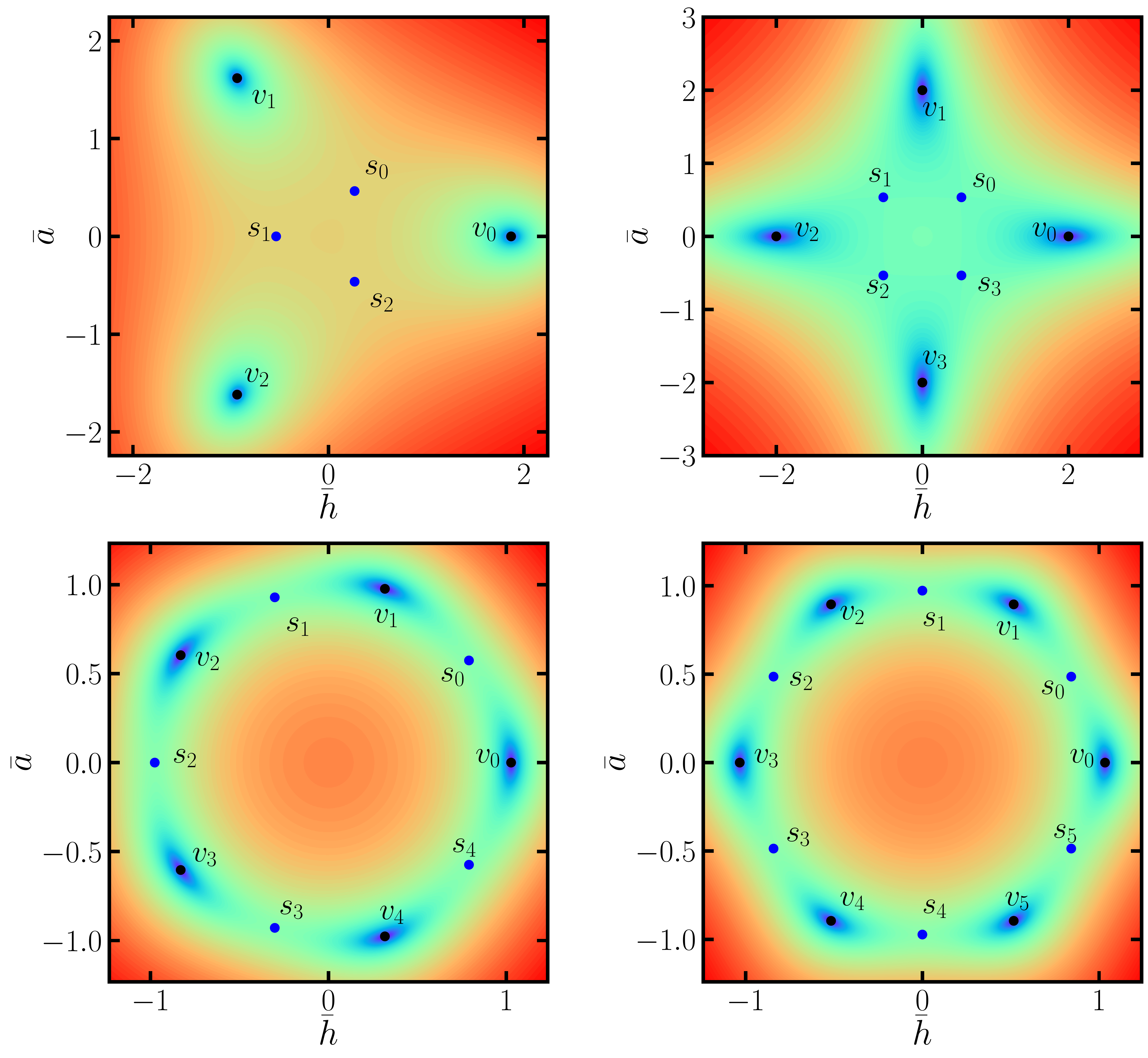}
\caption{Potential and vacuum properties in $Z_3$ (top-left), $Z_4$ (top-right), $Z_5$ (bottom-left) and $Z_6$ (bottom-right) symmetries. $v_i$ and $s_i$ are vacua and saddle points, respectively. The general form of the potential is given in Eq.~\eqref{eq:potential_Z_N} and the special cases with $N=3,4$ are given in Eqs.~\eqref{eq:potential_Z_3} and \eqref{eq:potential_Z_4}. $\lambda_2 =\frac{4\sqrt{2}}{9} \lambda_1^{1/2}$ in $Z_3$, $\lambda_2 =\frac{3}{8} \lambda_1$ in $Z_4$, $\lambda_2 =\frac{\sqrt{2}}{50} \lambda_1^{3/2}$ in $Z_5$ and $\lambda_2 =\frac{1}{25} \lambda_1^2 $ in $Z_6$ are used. $\mu/\sqrt{\lambda_1}$ is normalised to 1 in all panels.}
\label{fig:potential_ZN}
\end{figure*}

We give some examples of the potential form.
For $N=3$ and $4$, we consider renormalisable potential forms. Namely, only polynomials of $\phi^* \phi$ and $\phi^N + \phi^{* N}$ and the power of scalar fields in all terms no more than 4 are considered. They are given by
\begin{eqnarray}
V_{Z_3} &=&  -\mu^2 \phi^* \phi + \lambda_1 (\phi^* \phi)^2 - \lambda_2 \mu (\phi^3+\phi^{*3})\,, \label{eq:potential_Z_3}\\
V_{Z_4} &=&  -\mu^2 \phi^* \phi + \lambda_1 (\phi^* \phi)^2 - \lambda_2 (\phi^4+\phi^{*4})\,. \label{eq:potential_Z_4}
\end{eqnarray}
where all coefficients are real, the dimensional parameter $\mu>0$ is assumed without loss of generality, and $\lambda_1>0$ for $N=3$ ($\lambda_1 >|\lambda_2|$ for $N=4$) and the minus sign in front of $\lambda_2$ will be convenient to keep a positive $v_0$. These potentials are the most general $Z_3$- and $Z_4$-invariant potential which can gain non-trivial and stable vacua, respectively. Their dependencies on the scalar are shown in the top panel of Fig.~\ref{fig:potential_ZN}. We indicate vacua $v_i$ and saddle points $s_i$ of the potential (for $i=1,2, \cdots, N-1$) in the figure. At $N=3$, all degenerate vacua are adjacent in the field space. At $N=4$, vacua may not be adjacent with each other, e.g., $v_0$ and $v_2$. This case leads to different domain walls properties as will be detailed discussed in the next section.

For larger $Z_N$ with $N  \geqslant 5$, the renormalization requirement forbids terms such as $\phi^N$. Therefore, the theory is accidentally enlarged into $U(1)$. To impose $Z_N$ instead of $U(1)$ to be the symmetry of a theory, one approach is to treat the potential as an effective theory. In this case, we can simply write out the potential in the form
\begin{eqnarray} \label{eq:potential_Z_N}
V_{Z_N}^{\rm eff} &=&  -\mu^2 \phi^* \phi + \lambda_1 (\phi^* \phi)^2 - \lambda_2 \mu^{4-N} (\phi^N+\phi^{*N}) \,,
\end{eqnarray}
where only the leading higher-dimensional operator is considered as seen as the third term on the right hand side. This term is in general much smaller than the first two terms as it is a higher-order correction. In this case, an approximate $U(1)$ symmetry is preserved before the SSB.
The potential, up to an irrelevant constant term, is approximately re-written in the form
\begin{eqnarray} \label{eq:potential_Z_N_2}
V_{Z_N}^{\rm eff} \approx \lambda_1 (|\phi|^2 - v_0^2)^2 + 2 \lambda_2 \mu^{4-N} v_0^N [1 - \cos (N \theta)]\,,
\end{eqnarray}
where in this particular case $v_0 \sim \sqrt{\mu^2/(2\lambda_1)}$. The examples for $N=3$, 4, 5 and 6 are given in Fig.~\ref{fig:potential_ZN}.

Potentials as in Eqs.~\eqref{eq:potential_Z_3}--\eqref{eq:potential_Z_N} include only one term $\phi^N + \phi^{*N}$ which breaks the $U(1)$ symmetry. For this kind of potential, the CP symmetry is accidental. Given the more general from of potential with Hermitean conditions satisfied, we have,
\begin{eqnarray}
 -\mu^2 \phi^* \phi + \lambda_1 (\phi^* \phi)^2 - \lambda_2 \mu^{4-N} (e^{i\alpha}\phi^N+e^{-i\alpha}\phi^{*N})\,,
\end{eqnarray}
where the same conditions for $\mu$ and $\lambda_1$, $\lambda_2$ are satisfied as before and $\alpha$ is an arbitary phase. With the phase rotation $\phi \to e^{i\alpha/N} \phi$, we arrive at Eqs.~\eqref{eq:potential_Z_3}--\eqref{eq:potential_Z_N}.\footnote{This phase rotation may induce an explicit CP violation in the coupling for $\phi$ with other particles. } Furthermore, one can keep $\lambda_2$ always positive from the phase rotation. In the following discussion, we always keep all coefficients $\mu$, $\lambda_1$ and $\lambda_2$ positive in these equations.

We further mention a special case, a complex scalar in $Z_2$. Imposing a CP symmetry, we obtain $T^2=S^2=1$ and $TS=ST$. The whole symmetry is enlarged into the Klein symmetry $Z_2 \times Z_2^{CP}$. There are two degenerate vacua, both are real and connected by $Z_2$ transformation $v_1= -v_0$. In the domain wall solution, one can always fix the imaginary part $a=0$, and thus, the real component $h$ keeps similar behaviour as the real scalar in $Z_2$. We will not extend the discussion. However, note that by  suitably arranging the potential terms, spontaneous CP violation can be achieved. It generates CP domain walls with properties different from the classical $Z_2$ domain walls. Special examples of CP domain walls have been discussed in the context of extension of the SM \cite{Chen:2020wvu,Chen:2020soj}. For general properties of CP domain walls, we refer to our upcoming work.  In the rest of this paper, we will concentrate on domain walls with only CP conservation.

\subsection{Potential with more scalars} \label{sec:more_scalars}

To construct a UV-complete  $Z_N$-invariant theory with $N \geqslant 5$, more scalars have to be included. We include one more scalar, labelled as $\xi$ below. Charges of $\phi$ and $\xi$ in $Z_N$ are denoted as $q_\phi$ and $q_\xi$. It is hard to make a general statement as there could be many different ways of charged assignments in $Z_N$-invariant theory. However, we could specify several representative examples  which may result in different domain wall structures after the symmetry breaking.
Below are three categories we will discuss in the paper:
\begin{itemize}
\item[C1)] Charges of  $\phi$ and $\xi$ are coprime with $N$, i.e., ${\rm gcd} (q_\xi, N) = {\rm gcd} (q_\phi, N) = 1$.

\item[C2)] $q_\xi$ has a non-trivial common divisor of $N$, but $q_\phi$ is still coprime with $N$, i.e., ${\rm gcd} (q_\xi, N) > 1$ and ${\rm gcd} (q_\phi, N) =1$.

\item[C3)] Both $q_\phi$ and $q_\xi$ have non-trivial common divisors with $N$, i.e., ${\rm gcd}(q_\phi, N)$, ${\rm gcd}(q_\xi, N) >1$. We further require these two gcds are coprime with each other without loss of generality, otherwise, the essential symmetry is not $Z_N$ but $Z_{N/{\rm gcd}({\rm gcd}(q_\phi, N), {\rm gcd}(q_\xi, N))}$.

\end{itemize}

For each category, we consider a special example of potential with the following $Z_N$ and scalar charges arrangements.
For the first category, we take the $Z_5$ case as a typical example, with $\{N, q_\phi, q_\xi\} = \{5, 1, 2\}$. In this example, the general $Z_5$-invariant potential is given by
\begin{eqnarray} \label{eq:V_C1}
V_{Z_5}^{\rm C1} &=& -\mu^2 \phi^* \phi + \lambda_1 (\phi^* \phi)^2 -\mu_\xi^2 \xi^* \xi + \lambda_\xi (\xi^* \xi)^2 \nonumber\\
&& - \lambda_{\phi \xi 1} (\phi^3 \xi+\phi^{*3} \xi^{*})
- \lambda_{\phi \xi 2} (\phi \xi^{*3}+\phi^{*} \xi^{3})  \nonumber\\
&& - \lambda_{\phi \xi 3} \mu (\phi^2 \xi^*+\phi^{*2} \xi)
 - \lambda_{\phi \xi 4} \mu (\phi \xi^2+\phi^{*} \xi^{*2})\,.
\end{eqnarray}
Note that necessary conditions among coefficients in the potential are required, which will not be repeated below.
For the second and third categories, we take $N=6$. Charges are arranged as follows, $\{N, q_\phi, q_\xi\} = \{6, 1, 3\}$ in C2) and $\{N, q_\phi, q_\xi\} = \{6, 2, 3\}$ in C3).  In an economical consideration, $\xi$ can be assumed as a real scalar. Renormalisable potentials in these examples are written as
\begin{eqnarray} \label{eq:V_C2}
V_{Z_6}^{\rm C2} &=& -\mu^2 \phi^* \phi + \lambda_1 (\phi^* \phi)^2 - \frac{1}{2}\mu_\xi^2 \xi^2 + \frac{1}{4}\lambda_\xi \xi^4 \nonumber\\
 &&- \lambda_{\phi \xi} (\phi^3+\phi^{*3}) \xi \,,\\
\label{eq:V_C3}
V_{Z_6}^{\rm C3} &=& -\mu^2 \phi^* \phi + \lambda_1 (\phi^* \phi)^2 - \lambda_2 (\phi^3+ \phi^{*3})\nonumber\\
&& - \frac{1}{2}\mu_\xi^2 \xi^2 + \frac{1}{4}\lambda_\xi \xi^4 \,.
\end{eqnarray}
The last example gives decoupled scalar potential terms between $\phi$ and $\xi$.

In all these cases, our goal is to explore new features of domain walls which may be distinguished from those from a single scalar case. We will consider only the hierarchical scenario that $\xi$ decouples earlier than $\phi$. It  is  further worthy to restrict our discussion in the region $\mu_\xi^2 > 0$. For negative $\mu^2_\xi$, $\xi$ has a heavy mass $m_\xi \sim \sqrt{-\mu^2_\xi}$ before it gains a vacuum expectation value (VEV). After the scale drops below its mass scale, $\xi$ decouples and we arrive at an effective potential of a single scalar as in Eq.~\eqref{eq:potential_Z_N}. Therefore, a negative $\mu^2_\xi$ gives no distinguishable features of domain walls in a single scalar case.

\subsection{Applications in physical systems}

$Z_N$-invariant theories have been introduced as new physics. While $Z_2$ has a lot of applications, e.g., the R-parity symmetry in supersymmetric models, left-right parity in left-right symmetric model and its GUT extensions, those for $N \geqslant 3$ have been widely considered in several different frameworks. Below we show a few examples for these applications with $N \geqslant 3$.

One widely studied type including $Z_N$ symmetries is the axion model. QCD axions are proposed to solve the strong CP problem~\cite{Peccei:1977hh,Weinberg:1977ma,Wilczek:1977pj,Kim:1979if,Shifman:1979if,Dine:1981rt,Zhitnitsky:1980tq}, other mechanisms also predicts axion-like particles~\cite{Svrcek:2006yi,Arvanitaki:2009fg,Acharya:2010zx,Dine:2010cr} (see~\cite{Marsh:2015xka} for a recent review). In the axion framework, a global  $U(1)$ Peccei-Quinn (PQ) symmetry~\cite{Peccei:1977hh} is introduced  accompanied with a complex scalar $\phi = |\phi| e^{i\theta}$. $|\phi|$ gains a VEV $v_0$ at very high scale, breaking $U(1)$ spontaneously. Below the $U(1)$ breaking scale,  the theory is invariant under the PQ phase rotation, $\theta \to \theta + {\rm const}$ at classical level. However, at quantum level, it is broken due to the chiral anomaly for PQ-charged quarks. The latter induces a term $\sim N\theta G \tilde{G}/32\pi^2$, where $N$ appears as an integer combination of PQ charges of quarks. As a consequence, a small effective term for the phase $\theta$ is effectively generated
\begin{eqnarray}
V \supset \frac{m_a^2 v_0^2}{N^2} [1 - \cos (N \theta)]\,,
\end{eqnarray}
where $m_a$ is the axion mass.
This is consistent with the approximation in Eq.~\eqref{eq:potential_Z_N_2}. It satisfies the shift symmetry $\theta \to \theta + 2 \pi k/N$ for $k=0,1,2,..., N-1$. Therefore, a $Z_N$ is left unbroken as a residual symmetry after $U(1)$ SSB. Domain walls arise from axion-like framework have been well studied \cite{Vilenkin:1982ks,Hiramatsu:2012sc}. In the next section, we will not focus on walls of this form but treat it as a comparison to other walls from potentials with large $Z_N$ terms present.

Another important application of $Z_N$ is in addressing the fermion flavour puzzles. These puzzles refer to a series theoretical problems, including but not limited to understanding the highly hierarchical structure of charged fermion masses and origins of different mixing patterns in quark and lepton sectors. Given a $Z_N$ as the horizontal symmetry in the flavour space, $Z_N$-charged scalars, usually called flavons, become essential to generated flavoured fermions masses. They gains VEVs, breaking $Z_N$ spontaneously and generating flavour structures for  quark and lepton Yukawa couplings. Below we show two examples to see how an Abelian discrete symmetry solves these problems, one with a single scalar and the other with multiple scalars. The first example achieves the hierarchy fermion masses following the Froggatt-Nielsen mechanism \cite{Froggatt:1978nt},
and the second one is helpful to realise two-zero flavour textures in the fermion mass matrix \cite{Grimus:2004hf}.

The Froggatt-Nielsen mechanism was originally proposed with a global $U(1)$ symmetry \cite{Froggatt:1978nt}, and switching it to $Z_N$ is straightforward. The latter has been widely used in flavour model construction, in particular in leptonic flavour models when complimentary to a non-Abelian flavour symmetry (see \cite{King:2017guk,Morozumi:2017rrg,Feruglio:2019ktm} for recent reviews). In the following, we give a toy model with a scalar flavon $\phi$ in $Z_N$ with $N$ not specified. By assigning $\phi$ a unit charge and fermion charges $q_{F_{\alpha}}$ and $q_{f_{\beta}}$, Yukawa couplings of fermions are replaced by higher-dimensional operators involving $\phi$,
\begin{eqnarray} \label{eq:}
{\cal L}_Y = \sum_{\alpha \beta} \lambda_{\alpha \beta} \left( \frac{\phi}{\Lambda}\right)^{n_{\alpha\beta}} \overline{F}_{\alpha}  \overset{(\sim)}{H} f_{\beta} + {\rm h.c.},
\end{eqnarray}
where $F_{\alpha}$ is the SM electroweak doublet fermions and $f_{\alpha}$ is the right-handed fermion singlet with flavour indices $\alpha$ and $\beta$ respectively. $H$ is the SM Higgs, and it does not have to take a $Z_N$ charge here, $\overset{\sim}{H} = i\sigma_2 H^*$. For the left-handed doublet $F = Q \equiv \begin{pmatrix}u \\d \end{pmatrix}_L$ and $L\equiv \begin{pmatrix}\nu \\ l \end{pmatrix}_L$, the right-handed singlet $f = u_R, d_R$ and $\nu_R, l_R$, respectively. $n_{\alpha\beta}$ is an integer required by the $Z_N$ invariance, $n_{\alpha\beta} = q_{F_{\alpha}} - q_{f_{\beta}} ~({\rm mod}~ N)$.
 Yukawa couplings are effective consequences after the scalar $\phi$ gains the VEV,
\begin{eqnarray}
(Y_f)_{\alpha \beta} = \lambda_{\alpha \beta} \epsilon^{n_{\alpha \beta}},
\end{eqnarray}
where $\epsilon = \langle \phi \rangle / \Lambda$.
By arranging different charges for different flavours, each entry of the Yukawa coupling matrix is suppressed by $\epsilon^{n_{\alpha \beta}}$ with a flavour-dependent integer $n_{\alpha \beta}$. In this toy model, we have ignored couplings between $\phi^*$ and $\overline{F}_{\alpha} H f_{\beta}$. These terms should be either included in a complete model or forbidden by imposing additional symmetries.

\begin{table}
\begin{center}
\begin{tabular}{| l | c c c | c c c | c c c | c |}
\hline
particles & $\phi_1$ & $\phi_2$ & $\phi_3$ & $F_{1}$ & $F_{2}$ & $F_{3}$ & $f_{1}$ & $f_{2}$ & $f_{3}$ & $H$ \\\hline
$Z_6$ charge & 3 & 2 & 1 & 0 & $-2$ & $-1$ & 0 & 2 & 1 & 1 \\\hline
\end{tabular}
\end{center}
\caption{An example of charges of scalars and fermions in $Z_6$ to achieve two-zero textures of fermion flavour structures.}
\label{tab:charge_2}
\end{table}

The texture-zero approach was developed to calculate the Cabibbo angle of quark flavor mixing \cite{Weinberg:1977hb,Wilczek:1977uh,Fritzsch:1977za}. It has been applied in both quark and lepton Yukawa couplings (see \cite{Xing:2019vks} for a recent review). The approach follows the idea that if some elements of the fermion mass matrices are vanishing, the number of free parameters will be reduced and there exist some testable relations between the fermion mass ratios and the observable flavour mixing quantities. A typical two-zero texture takes the form
\begin{eqnarray} \label{eq:two-zeros}
Y_f = \begin{pmatrix} 0 & C_f & 0 \\ C_f' & \tilde{B}_f & B_f \\ 0 & B_f' & A_f \end{pmatrix}.
\end{eqnarray}
This pattern now is still consistent with the updated flavour data in both quark and lepton sectors \cite{Fritzsch:2011qv, Zhou:2012ds, Xing:2015sva, Fritzsch:2021ipb}. While the original proposal of texture zeros is for the phenomenological interest on correlations of fermion masses and mixing angles, imposing an Abelian discrete symmetry gives an explanation of its origin \cite{Grimus:2004hf}. Here we list an example for how that in Eq.~\eqref{eq:two-zeros} is realised in $Z_6$. We includes three scalars $\phi_1$, $\phi_2$ and $\phi_3$. These particles, again, do not take electroweak charges. Charges of these particles and fermions in $Z_6$ are listed in Table~\ref{tab:charge_2}.
General couplings for fermion masses under gauge symmetries and the horizontal symmetry up to dimension 5 are given by
\begin{eqnarray}
{\cal L}_Y = \sum_{i; \alpha,\beta} (\lambda_f^{(i)})_{\alpha\beta} \frac{\phi_i}{\Lambda} \, \overline{F}_{\alpha} \overset{(\sim)}{H} f_{\beta} + {\rm h.c.},
\end{eqnarray}
where the SM Higgs does not take a $Z_6$ charge, $i$ runs for three copies of scalars, and $\alpha,\beta$ run for three flavours of fermions. All $Y_f^{(i)}$ for $i=1,2,3$ are $3\times 3$ matrices. As the theories is invariant $Z_6$, they have to follow the following forms
\begin{eqnarray}
&\lambda_f^{(1)} = \begin{pmatrix} 0 & \times & 0 \\ \times & 0 & 0 \\ 0 & 0 & \times \end{pmatrix}, \quad
\lambda_f^{(2)} = \begin{pmatrix} 0 & 0 & 0 \\ 0 & 0 & \times \\ 0 & \times & 0 \end{pmatrix}, \nonumber\\
&\lambda_f^{(3)} = \begin{pmatrix} 0 &0 & 0 \\ 0 & \times & 0 \\ 0 & 0 & 0 \end{pmatrix},
\end{eqnarray}
where a cross represents a non-vanishing entry. After $\phi_k$ gain VEVs, the Yukawa coupling matrix appears as a linear combination, i.e., $Y_f = \sum \lambda_f^{(i)} \epsilon_i$ with $\epsilon_i = \langle \phi_i \rangle / \Lambda$, and thus takes the form of Eq.~\eqref{eq:two-zeros}.
Note that the zero entries could gain corrections from higher-dimensional operators. These corrections can be suppressed by a natural arrangement $\epsilon \sim y_{\tau,b} \sim 0.01$. To generate an $\mathcal{O}(1)$ top Yukawa coupling, one can simply re-assign charges of $t_{R}$ such that the top-quark Yukawa coupling is generated from a renormalisable term. A way to fully forbid these corrections is considering a renormalisable variation of the model that the three scalars are replaced by three electroweak Higgs doublets (where the lightest one after mix gives the Standard Model Higgs) and directly arrange $Z_6$ charges for them. We refer to \cite{Zhou:2012ds} for more details of model building.

In these flavour models, $\epsilon$ or $\epsilon_i$ are usually correlated with fermion masses and mixing angles, which could be restricted by fitting flavour data. The VEV $\langle \phi \rangle$, or equivalently the SSB scale of $Z_N$, is undetermined, and usually assumed at a very high energy scale beyond the capability of direct searches in laboratory.



In addition, $Z_3$-invariant Next-to-Minimal Supersymmetric Standard Model (NMSSM) has used considered to forbid unnecessary couplings between the Higgs singlet and doublets in the superpotential~\cite{Ellwanger:2009dp}. In the NMSSM, the VEV of the singlet $S$ generates the $\mu$ term in the Higgs potential, helping to trigger the electroweak symmetry breaking (EWSB). Before the EWSB, the singlet potential can be written as
\begin{eqnarray}
V\supset m_S^2|S|^2+\left(\frac{1}{3}\kappa A_\kappa S^3+{\rm h.c.}\right)+|\kappa|^2|S|^4,
\end{eqnarray}
which can be matched to the $Z_3$ case in our discussion.

\section{Domain walls triggered by a complex scalar \label{sec:small_Z_N}}

Domain walls form after a discrete symmetry is broken. Given a scalar $\phi$ with a potential $V(\phi)$ which is invariant under transformations of a discrete symmetry, degenerate vacua $\langle \phi \rangle = v_0$, $v_1$, ... exist. A domain wall refers to a solution of the equation of motion (EOM) for the scalar $\phi$ along one spatial dimension
\begin{eqnarray}
\frac{d^2\phi}{d z^2} = \frac{\partial V(\phi)}{\partial \phi}
\end{eqnarray}
with different vacua on the two sides $z \to \pm \infty $. For boundary conditions
\begin{eqnarray}
\phi|_{z\to -\infty} = v_i\,,\quad
\phi|_{z\to +\infty} = v_j
\end{eqnarray}
with $i \neq j$, we denote the corresponding domain wall as $\boxed{v_i | v_j}$, and this notation will be very useful in $N\geqslant3$ cases. For $Z_N$ with $N=2,3,4$ renormalisable potentials of a single scalar is enough to breaking the symmetry. We call these symmetries as small $Z_N$ symmetries. In this section, we will discuss domain wall properties from these symmetries. Those from larger $Z_N$ symmetries will be discussed in the next section.

\subsection{Domain walls from $Z_2$ breaking}

The simplest type is a $Z_2$ domain wall. Given a real scalar or a complex scalar with real component with potential in Eq.~\eqref{eq:potential_Z2} the $Z_2$ domain wall solution of $\phi$ along $z$ is
\begin{eqnarray}\label{Z2_solution}
\phi = v_0 \tanh\left( \sqrt{\frac{\lambda}{2}} v z \right) \,,
\end{eqnarray}
which can be denoted as $\boxed{v_0|v_1}$ in our notation with $v_1=-v_0$. There are two important parameters for domain walls. The first one is the tension of the wall, which measures the energy stored per unit area on the wall. The second one is thickness of the wall, which estimates the typical length scale of the scale variation of $\phi(z)$.

The tension of the wall is calculated via the energy momentum tensor, $T_{\mu\nu} =  \partial_\mu \phi^* \partial_\nu \phi - {\cal L} g_{\mu\nu}$. Along the direction perpendicular to the wall, the $(0,0)$ entry, i.e. the energy density component, is
\begin{eqnarray}
\varepsilon (z) \equiv T_{00}= \frac{1}{2} \left[\frac{d\phi(z)}{dz}\right]^2 + \Delta V(\phi(z))\,,
\end{eqnarray}
where $\Delta V(\phi) = V(\phi) - V_{\rm min}$.
The integration along $z$ gives
\begin{eqnarray}
\sigma = \int_{-\infty}^{\infty} d z \varepsilon(z)\,.
\end{eqnarray}
This is also called the tension of the wall. In the $Z_2$ case, this property is calculated to be $\sigma = \frac{4}{3} \sqrt{\frac{\lambda}{2}} v^3$.

The thickness of domain walls in the classical $Z_2$ case is defined as the factor appearing in the a hyperbolic tangent function of the scalar profile $ \propto \tanh(z/\delta)$, i.e., $\delta \approx \sqrt{2/(\lambda v^2)}$ \cite{Vilenkin:1984ib}. This definition leads to
\begin{eqnarray} \label{eq:thickness_def}
\int_{-\delta/2}^{\delta/2} dz \varepsilon(z) \approx 64\% \times \sigma.
\end{eqnarray}
For domain walls beyond $Z_2$, scalar profiles do not hold a hyperbolic-tangent behaviour any more. However, we can apply Eq.~\eqref{eq:thickness_def} as a generalised definition of the wall thickness for all kinds of scalar profiles in domain wall solutions.

\subsection{Domain walls from $Z_3$ breaking}

The $Z_3$ domain wall has been solved explicitly in Ref. \cite{Wu:2022stu}. Here, we give a summary of its main property. By fixing $\phi=v_0$ at $z=-\infty$ and $\phi=v_1$ at $z= +\infty$, we are able to obtain the $\boxed{v_0|v_1}$ domain wall solution by solving the EOM of $h$ and $a$ with potential given in Eq.~\eqref{eq:potential_Z_3}. It is convenient to introduce a dimensionless parameter $\beta \equiv 3\lambda_2/\sqrt{8\lambda_1}$. By performing a normalisation of the field and coordinate
\begin{eqnarray} \label{eq:normalisation}
\bar{h} = \frac{\sqrt{\lambda_1}}{\mu} h \,, \quad
\bar{a} = \frac{\sqrt{\lambda_1}}{\mu} a \,, \quad
\bar{z} = \mu \, z\,,
\end{eqnarray}
all variables are dimensionless and the system depends only $\beta$. This parameter represents how far a $Z_3$-invariant theory deviate from the $U(1)$ symmetry. In the limit $\beta \to 0$, a global $U(1)$ is recovered.

\begin{figure*}
\centering
\includegraphics[width=.9\textwidth]{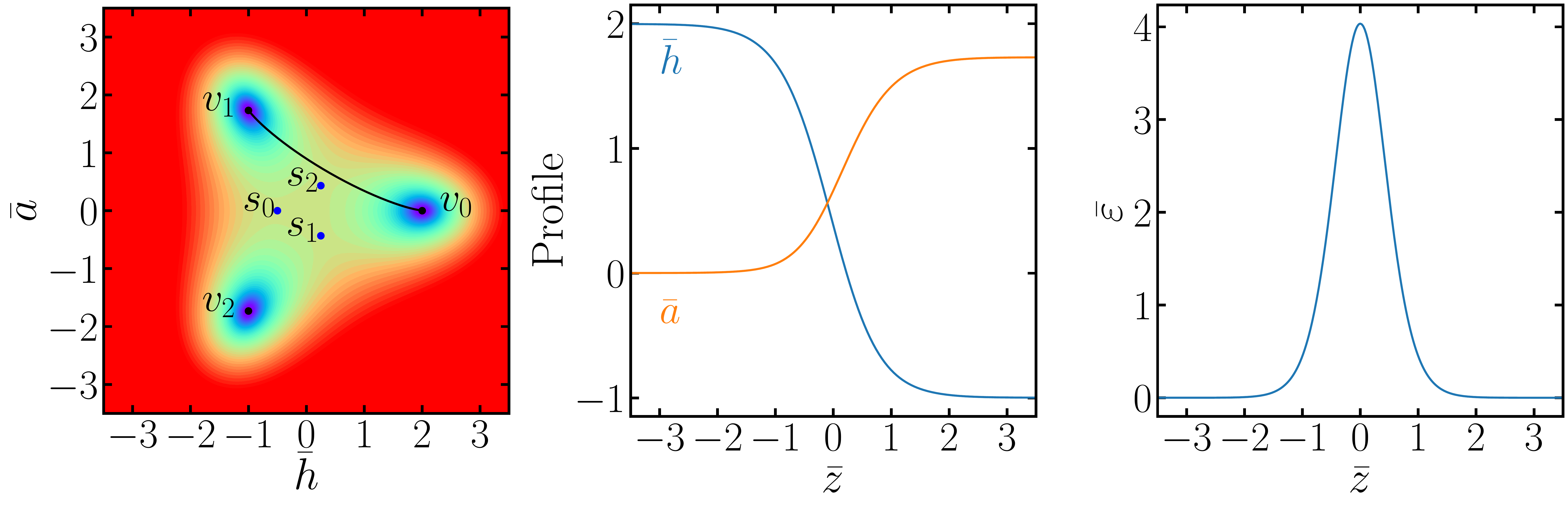}
\caption{Vacuum property and domain walls in $Z_3$ symmetry breaking. The left panel shows a contour plot of a $Z_3$-invariant potential in the scalar field space. Three degenerate vacua are indicated as $v_0$, $v_1$ and $v_3$ and three saddle points are labelled by $s_0$, $s_1$ and $s_2$. We obtain the solution for the domain wall $\boxed{v_0|v_1}$, i.e., the soliton solution with vacuum $v_0$ fixed at $z \to - \infty$ and $v_1$ at $z \to + \infty$. The path in the field space is noted as the solid curve in the left panel, field profiles along the coordinate $z$ are shown in the middle panel, and the field energy density along $z$ are given in the right panel. Fields and coordinate are normalised as in Eq.~\eqref{eq:normalisation}. $\beta \equiv 3\lambda_2/\sqrt{8\lambda_1} = 3/4$ are used.
Figure copied from Ref. \cite{Wu:2022stu}.}\label{fig:DW_Z3}
\end{figure*}

The EOM of $\bar{h}$ and $\bar{a}$ is numerical solved for fixed values of $\beta$. In Fig.~\ref{fig:DW_Z3}, we show the bubble wall solution at $\beta = 3/4$. The path in the complex plane of $\phi$ is shown in the left panel. The middle panel gives the scalar profiles as functions of the spatial coordinate $z$, which is perpendicular to the wall. The corresponding energy density stored in the scalar along $z$ is shown in the right panel. Here the energy density has been normalised
\begin{eqnarray}
\bar{\varepsilon} = \frac{1}{2} (\bar{h}^{\prime 2} + \bar{a}^{\prime 2}) +  \Delta \bar{V} \,,
\end{eqnarray}
and $\Delta \bar{V} =\Delta V \lambda_1/\mu^4 $.
The tension and thickness of the wall are derived to be
\begin{eqnarray} \label{eq:sigma}
\sigma =  \frac{\mu^3}{\lambda_1} \bar{\sigma}\,;  \quad
\delta = \frac{\bar{\delta}}{\mu}\,,
\end{eqnarray}
where $\bar{\sigma}$ and $\bar{\delta}$ are the normalised tension and thickness defined via
\begin{eqnarray}
\bar{\sigma} = \int_{-\infty}^{\infty} d \bar{z} \bar{\varepsilon}(\bar{z}) \,;\quad
\int_{-\bar{\delta}/2}^{\bar{\delta}/2} d \bar{z} \bar{\varepsilon}(\bar{z}) = 64\% \times \bar{\sigma},
\end{eqnarray}
respectively. As $\bar{\sigma}$ and $\bar{\delta}$ depend on only one free parameter $\beta$, we can calculate them by scanning $\beta$ in a wide range. Dependences of $\bar{\sigma}$ and $\bar{\delta}$ on $\beta$ can be fitted with semi-analytical formulas.
\begin{eqnarray}
\bar{\sigma}(\beta) &=& 2.18 \beta^{0.5}+ 1.8 \beta^{1.85} + 4 \beta^3
 \,, \nonumber\\
\bar{\delta}(\beta) &=& 0.64 \beta^{-0.5} \frac{1+ 3.07 \beta^{1.37}}{1 + 0.607 \beta^{0.5} + 1.86 \beta^{1.87}} \,.
\end{eqnarray}
These formulae match with the numerical results very well with relative errors less than $2\%$ for $10^{-3} \leqslant \beta \leqslant 10^4$. Expressing $\mu$ and $\lambda_1$ in terms of physical observables, $\sigma$ and $\delta$ can be re-expressed as $\sigma = m_a v_0^2 f(\beta)$ and $\delta = m_a^{-1} g(\beta)$, where $f(\beta)$ and $g(\beta)$ are both order-one functions with expressions given in Eqs.~(14) and (17) of Ref. \cite{Wu:2022stu}, respectively.

In the case $\beta \ll 1$,  i.e., $\lambda_2 \ll \sqrt{\lambda_1}$, a global $U(1)$ is approximately conserved at high energy scale. We encounter a two-step SSB as the temperature decreases during the Hubble expansion.
The first step is the SSB of $U(1)$, which happens around the scale $v_0 \simeq \mu/\sqrt{2\lambda_1}$.
A well-known consequence following the $U(1)$ breaking is the formation of cosmic strings with the string tension $\sim \pi v_0^2$ \cite{Hindmarsh:1994re}. The SSB of $U(1)$ leads to a pseudo-Nambu-Goldstone boson with mass $m_a^2 \simeq 3 \beta m_h^2 \ll m_h^2$.
The next step is the SSB of $Z_3$. It happens when the temperature decreases into the energy scale comparable with $\sqrt{m_a M_{\rm P}}$ with $M_{\rm P}$ the Planck mass. Around this energy scale,  the third term of Eq.~\eqref{eq:potential_Z_3} becomes non-negligible with the gradient energy, which is of order $v_0^2 H^2$ and $H$ the Hubble parameter. After the SSB of $Z_3$, the phase of $\phi$'s VEV is then fixed to one of the three phases $0$, $2\pi/3$ and $4\pi/3$. The energy barrier between spatial regions with different phases soon forms a domain wall with tension $\sigma \simeq 2.18 \sqrt{\beta} \mu^3/\lambda_1 \simeq 1.8 m_a v_0^2$. Thus, we arrive at a topological defect of walls bounded by strings  \cite{Kibble:1982dd, Everett:1982nm}. This object is just like a {\it revolving door} that each door on its boundary attaches to the axis in the centre.

\subsection{Domain walls from $Z_4$ breaking}

We first analyse the vacuum properties of the $Z_4$-invariant theories. The general renormalisable tree-level potential is given in \eqref{eq:potential_Z_4}, which includes only three terms $\phi^* \phi$, $(\phi^* \phi)^2$ and $\phi^4+\phi^{*4}$.  All parameters are real and positive without loss of generality. To ensure the potential to have a non-trivial stable vacuum,  $\lambda_1> 2 \lambda_2$ has to be satisfied. It is useful to parametrise $\beta = 2 \lambda_2/\lambda_1$ with $0<\beta<1$. Similar to the $Z_3$ case, $\beta$ here also characterises the level of deviation of the $Z_4$-invariant potential from the global $U(1)$ symmetry.

There are four solutions for the vacuum, classified as
\begin{eqnarray} \label{eq:vev}
v_k = \frac{\mu}{\sqrt{2\lambda_1(1-\beta)}} e^{i \frac{2 \pi}{4} k}\,,
\end{eqnarray}
for $k=0$, 1, 2, 3, as well as four saddle points
\begin{eqnarray}
s_k = \frac{\mu}{\sqrt{2\lambda_1(1+\beta)}} e^{i \pi (2k+1)/4} \,.
\end{eqnarray}
These solutions have been geometrically shown in the top-right panel of Fig.~\ref{fig:potential_ZN} in section~\ref{sec:potential}.

Profiles of the scalars from one vacuum tunneling to another is obtained by solving the EOM of $h$ and $a$, which is given by
\begin{eqnarray} \label{eq:EOM_Z4}
\bar{h}''(\bar{z}) &=& [-1+ \bar{h}^2+\bar{a}^2 + \beta (3\bar{a}^2-\bar{h}^2)] \bar{h} \,,\nonumber\\
\bar{a}''(\bar{z}) &=& [-1+ \bar{h}^2+\bar{a}^2 + \beta (3\bar{h}^2-\bar{a}^2)] \bar{a} \,,
\end{eqnarray}
where $\bar{h}$, $\bar{a}$ and $\bar{z}$ are normalised fields and coordinate defined in Eq.~\eqref{eq:normalisation}.
For the boundary conditions, however, we have to distinguish them into two branches.
\begin{itemize}
\item Two vacua are adjacent in the field space, e.g., $v_0$ and $v_1$, walls denoted as $\boxed{v_0|v_1}$.
\item Two vacua are non-adjacent in the field space, e.g., $v_0$ and $v_2$, walls denoted as $\boxed{v_0|v_2}$.
\end{itemize}
Note that the second branch has not been discussed in the study of classical $Z_2$ domain wall  or $Z_3$.
For walls separating adjacent and non-adjacent vacua, we denote them as adjacency walls and non-adjacency walls, respectively. We discuss them in the following.

\begin{figure*}
\centering
\includegraphics[width=.9\textwidth]{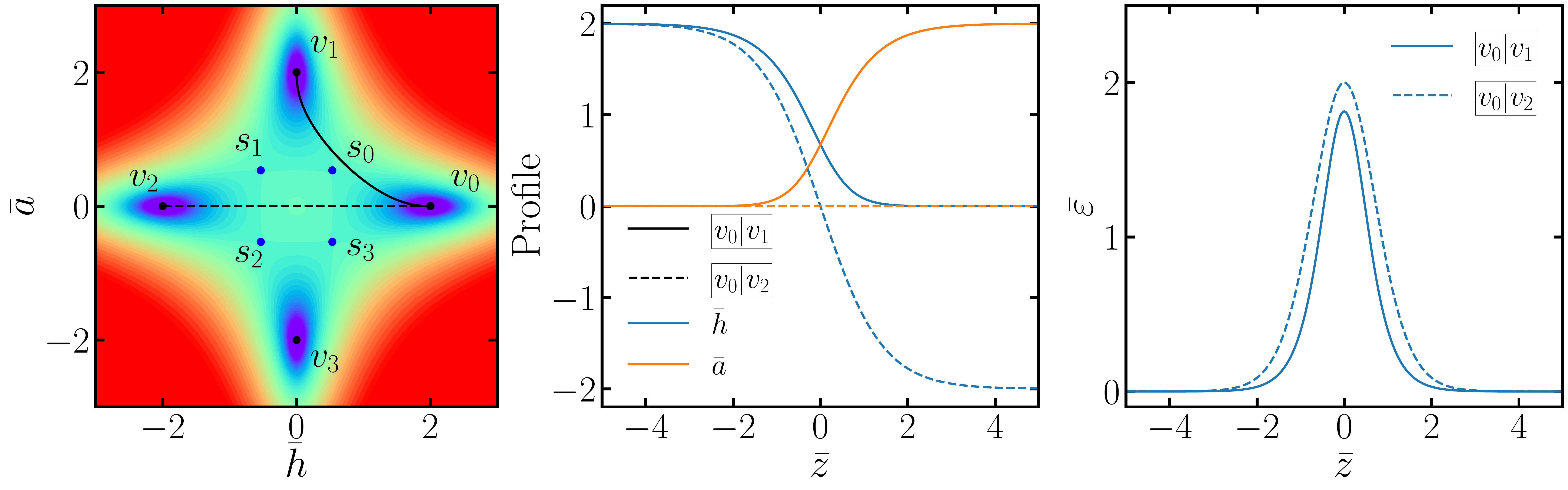}
\caption{Vacuum property and domain walls in $Z_4$ symmetry breaking. The solid and dashed curves refer to adjacency wall $\boxed{v_0|v_1}$ and non-adjacency wall $\boxed{v_0|v_2}$, respectively. $\beta \equiv 2 \lambda_2 / \lambda_1=3/4$ is used. Other conventions are the same as in Fig.~\ref{fig:DW_Z3}.}\label{fig:DW_Z4}
\end{figure*}

\begin{figure*}
\centering
\includegraphics[width=0.7\textwidth]{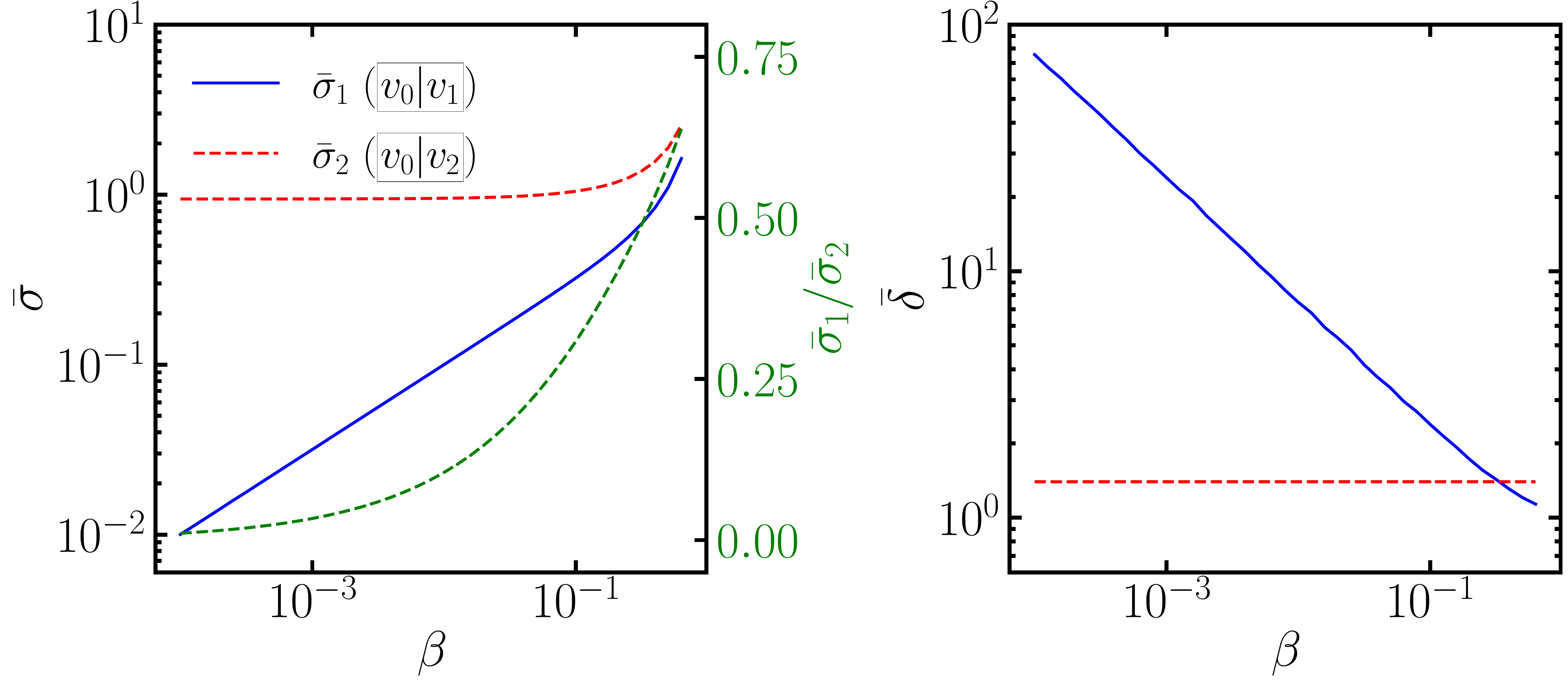}
\caption{Dependence of tension and thickness on $\beta$ for two kinds of $Z_4$ domain walls.}\label{fig:Z4_sigma_width}
\end{figure*}

Given the path from $v_0$ to $v_1$ for $z$ from $-\infty$ to $+\infty$, the solution of an adjacency wall $\boxed{v_0|v_1}$ is determined. The profiles of $h$ and $a$ along $z$ are given in the middle panel of Fig.~\ref{fig:DW_Z4}. And $\varepsilon$ along $z$ is given in the right panel of Fig.~\ref{fig:DW_Z4}.
Varying $\beta$, we obtain the domain wall tension $\sigma$ and the thickness $\delta$ as functions of $\beta$, as seen in Fig.~\ref{fig:Z4_sigma_width}. The solutions are semi-analytically given by
\begin{eqnarray} \label{eq:tension_Z4}
\bar{\sigma} (\beta) &=& 0.67 \beta^{0.5} \left( 1+ \frac{0.5}{1+4 \beta} \right)\,, \\
\bar{\delta} (\beta) &=& 0.75 \beta^{-0.5} + 0.33 \beta^{1.2}
\end{eqnarray}
at errors less than $2\%$ for $10^{-4} \leqslant \beta \leqslant 1 - 10^{-4}$. The solution has a singularity at $\beta=1$. Discussion in the limit $\beta \to 1$ is given in Appendix~\ref{app:Z4}. In the limit $\beta \ll 1$, we encounter domain walls bounded by strings as discussed in the $Z_3$ case. We have checked that our result of the wall tension is consistent with those obtained in axion-like models.

The non-adjacency wall is more complicated. For $\beta > 1/3$, we have first confirmed a hyperbolic tangent solution for $h$ if $a=0$ is fixed, e.g. the wall solution for $\boxed{v_0|v_2}$. This is the same as the classical solution obtained in the $Z_2$ domain walls \cite{Vilenkin:1984ib}. Thus, the tension and thickness of the wall are given by $\sigma_{Z_2}=\frac{4}{3}\sqrt{\frac{\lambda}{2}}v^3$ and $\delta_{Z_2}=(\sqrt{\frac{\lambda}{2}} v)^{-1}$ \cite{Vilenkin:1984ib,Saikawa:2017hiv}. Here in our particular case, $\lambda$ is replaced by $\lambda = \lambda_1(1 -\beta)$. We derive the normalised tension and thickness of the wall as
\begin{eqnarray}
\bar{\sigma}_{Z_2} (\beta) &=& \frac{2\sqrt{2}}{3(1-\beta)}\,, \nonumber\\
\bar{\delta}_{Z_2} (\beta) &=& \sqrt{2} \,,
\end{eqnarray}
respectively. Note that this solution holds only in the range $1/3 < \beta <1$.

\begin{figure*}
\centering
\includegraphics[width=.9\textwidth]{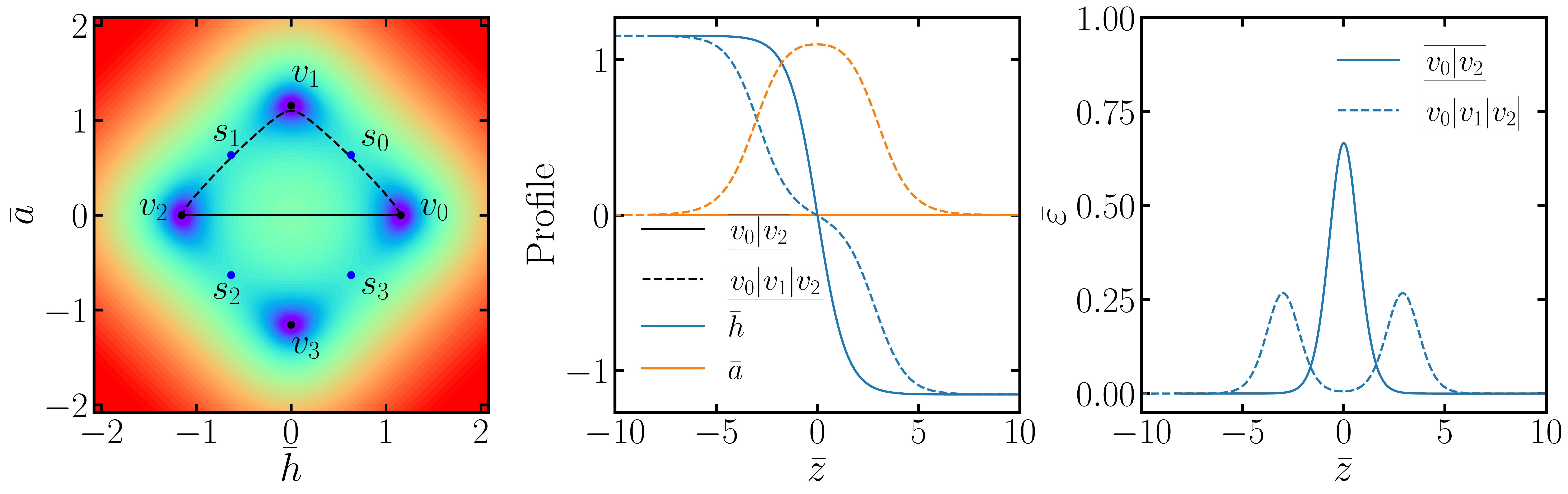}
\caption{Comparison of non-adjacency wall $\boxed{v_0|v_2}$ and two adjacency walls $\boxed{v_0|v_1|v_2}$ in $Z_4$ breaking. The same conventions are used as in Fig.~\ref{fig:DW_Z4} except $\beta = 1/4$.  The right panel shows that the tension of a non-adjacency wall is greater than the sum of tensions of two adjacency walls}\label{fig:DW_Z4_2}
\end{figure*}

For $0<\beta <1/3$, we observe another solution with the path and profile shown in dashed curves in Fig.~\ref{fig:DW_Z4_2}. This solution gives the path $v_0 \to v_1 \to v_2$ for $z$ from $-\infty \to 0 \to +\infty$, resulting in a domain wall $\boxed{v_0|v_1|v_2}$. At $\beta = 1/3$, $\bar{\sigma}_{Z_2} = \sqrt{2}$, which is twice of the tension calculated via Eq.~\eqref{eq:tension_Z4} (see Appendix~\ref{app:Z4} for an explicit proof). Once $\beta <1/3$, $\bar{\sigma}_{Z_2}$ is larger than the latter. Namely, the energy stored in the non-adjacency wall $\boxed{v_0|v_2}$ is larger than the total energy stored in the wall $\boxed{v_0|v_1|v_2}$. Thus, the non-adjacency wall becomes unstable. Even if it could form in some progresses, it will lose energy via splitting into two adjacency walls:
\begin{equation}
\boxed{v_0|v_2}\to\boxed{v_0|v_1|v_2}\to\boxed{v_0|v_1}+\boxed{v_1|v_2}\,.
\end{equation}
A new vacuum between the two adjacency walls, i.e., $v_1$ is generated after the wall splitting.
We have numerically checked that, for $\beta <1/3$ and given an initial path of the hyperbolic tangent solution with a small shift for $a$ from $0$, the path automatically deviates from the initial values during the deformation iteration and eventually stablised at the solution of the dashed curves in Fig.~\ref{fig:DW_Z4}. At the same time, the new vacuum $v_1$, separated by the two adjacency walls, is populated. Its volume in the three-dimensional space increases during the expansion of the universe until  the next stage when explicit-breaking terms dominate the evolution.

\subsection{Domain walls from larger $Z_N$ breaking}
We consider domain wall formation from SSB of $Z_N$ with $N \geqslant 5$.
For the single-scalar case, the effective potential is given in Eq.~\eqref{eq:potential_Z_N}. The $\phi^N$ term is a non-renormalisable operator, which in general should be sub-leading compared with the $U(1)$-invariant terms in the potential. We encounter two-step spontaneous symmetry breaking,
\begin{eqnarray}\label{eq:chain1}
{\rm Approx.}~U(1) \to Z_N \to 1 \,.
\end{eqnarray}
In the first step, $\phi$ gains a VEV $v_\phi e^{i \theta}$ with an arbitrary phase $\theta$, leading to the spontaneous breaking of the approximate $U(1)$ symmetry. Topologically, the SSB of $U(1)$ leads to the production of cosmic strings. In the second step, $|\langle\phi\rangle|$ is slightly modified to be $v_0$, and more importantly, the phase of $\langle \phi \rangle$ is fixed at one of the $N$ values $2 \pi k / N$ for $k=0,1,\cdots, N-1$. Domain walls form, with the boundary touching to the strings. Eventually, we arrive at a topological picture of ``{\it revolving doors}'': on the axle is a string, bounded by domain wall doors.

\begin{figure*}
\centering
\includegraphics[width=.9\textwidth]{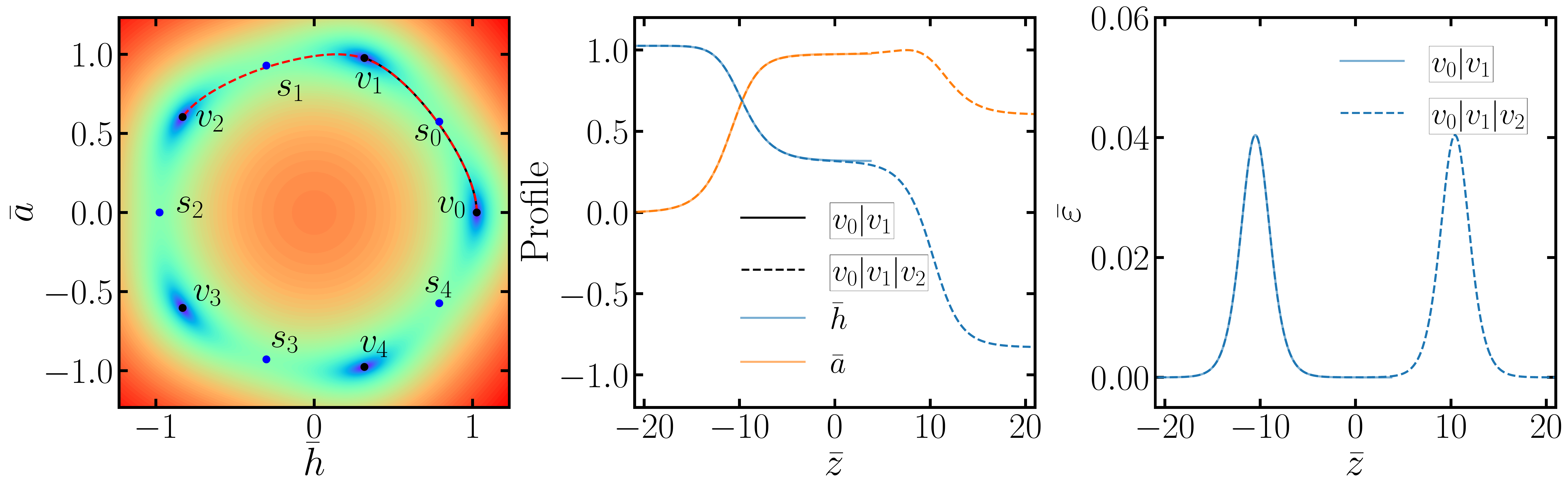}
\caption{Comparison between the domain wall solution for $\boxed{v_0|v_1}$ (solid curve) and that from $\boxed{v_0|v_1|v_2}$ (dashed curve) in $Z_5$ breaking. $\lambda_2 =\frac{\sqrt{2}}{50} \lambda_1^{3/2}$ is used and the rest conventions are the same as in Fig.~\ref{fig:DW_Z4}.}\label{fig:DW_Z5}
\end{figure*}

We have numerically checked that, as the $\phi^N$ term is small compared with the $U(1)$-invariant terms in the scalar potential, there is no non-adjacency wall solution. This is consistent with former results on axion-like potential, i.e., in the form of Eq.~\eqref{eq:potential_Z_N_2} \cite{Coleman:1985rnk,Vilenkin:2000jqa}. Namely, a solution from $v_2$ to $v_0$ always passes $v_1$. An example in $Z_5$ with potential  in Eq.~\eqref{eq:potential_Z_N} with $\lambda_2 =\frac{\sqrt{2}}{50} \lambda_1^{3/2}$ is given in Fig.~\ref{fig:DW_Z5}.

To end this section, we discuss the stability problem of these walls. Properties of walls in the whole paper are based on $Z_N$-invariant potentials.
It is well-known that explicit breaking is a necessary condition to solve the domain wall problem if the SSB of discrete symmetries happens below the inflationary scale.
The lifetime of $Z_2$ domain walls depends on the explicit breaking of the symmetry. The weaker the explicit breaking is, the longer domain walls leave. This statement is straightforwardly generalised to walls from SSB of $Z_3$ \cite{Wu:2022stu} and is expected to work for adjacent walls for larger $Z_N$.  The explicit breaking, as a totally independent term, can be assumed within a suitable range that the non-adjacent walls collapse before the BBN epoch, such that the observation does not conflict with the standard cosmology.
For the scalar below the SSB and above the wall-collapsing scale, we can treat these walls as stable topological defects.
However, non-adjacent walls in larger $Z_N$ could be unstable with small $Z_N$ effect even no explicit breaking is considered. As a consequence, non-adjacent walls bounded by strings are formed. In the $Z_4$ case, the size of $Z_4$ effect is characterised by a parameter $\beta$, and the wall become unstable if $\beta <1/3$. 

\section{Domain walls triggered by multiple scalars}\label{sec:large_Z_N}

In a renormalisable theory, one complex scalar is not enough to achieve the SSB of $Z_N$ for $N \geqslant 5$. A UV-complete theory requires more scalars. Once more scalars are involved, a general discussion on domain walls becomes impossible due to the large freedom of charge assignments of the scalars. In this section, we will focus on only two scalars $\xi$ and $\phi$ with charge assignments in three categories C1), C2) and C3) introduced in section~\ref{sec:more_scalars}. Following the discussions there, we assume $\xi$ gains a VEV to trigger the first step of SSB and $\phi$ leads to the second step of SSB. Examples of $Z_5$-invariant potential in Eqs.~\eqref{eq:V_C1} and $Z_6$-invariant potentials in Eqs.~\eqref{eq:V_C2} and \eqref{eq:V_C3} will be considered as case studies in the following two subsections. We then generalise the results in the last subsection.

\subsection{Domain walls from $Z_5$ breaking}

In C1), an example of $Z_5$-invariant potential is constructed with charges of $\phi$ and $\xi$ given by $1$ and $2$, respectively, as given in Eq.~(\ref{eq:V_C1}):
\begin{eqnarray*}
V_{Z_5}^{\rm C1} &=& -\mu^2 \phi^* \phi + \lambda_1 (\phi^* \phi)^2 -\mu_\xi^2 \xi^* \xi + \lambda_\xi (\xi^* \xi)^2 \\
&&- \lambda_{\phi \xi 1} (\phi^3 \xi+\phi^{*3} \xi^{*})
- \lambda_{\phi \xi 2} (\phi \xi^{*3}+\phi^{*} \xi^{3})  \\
&& - \lambda_{\phi \xi 3} \mu (\phi^2 \xi^*+\phi^{*2} \xi) - \lambda_{\phi \xi 4} \mu (\phi \xi^2+\phi^{*} \xi^{*2})\,.
\end{eqnarray*}
Two energy scales are important for this kind of potential. They are the $Z_5$ breaking scale denoted by the VEV $v_0$ and the new scalar $\xi$ mass scale $m_\xi$. We discuss the domain wall formation in the following scenario where a hierarchy between scales is satisfied, $m_\xi \gg v_0$. $U(1)$ is an approximately good symmetry at high scale. We again arrive at a two-step SSB, in which $U(1)$ breaking first and then $Z_5$, similar to the breaking chain in Eq.~\eqref{eq:chain1} in the single scalar case.

However, the breaking history is totally different from that in the single scalar case. For positive $\mu^2_\xi$, renormalisable terms of $\xi$, not $\phi$, trigger the spontaneous breaking of $U(1)$ directly. $\xi$ gains a VEV $v_\xi e^{i\theta_\xi}$ with the absolute value $v_\xi \sim \sqrt{\mu_\xi^2/2\lambda_\xi}$ and an arbitrary phase $\theta_\xi$. It leads to the spontaneous breaking of the $U(1)$ symmetry which is approximately preserved at high scale. The radial component along the VEV of $\xi$ gains a mass $m_\xi \sim \sqrt{2\mu_\xi^2}$,  leading to the effectively potential dominated by
\begin{eqnarray}
V_{Z_5}^{\rm eff} &=& -\mu^2 |\phi|^2 + \lambda_1 |\phi|^4 - \lambda_{\phi \xi 1} |\phi|^3 v_\xi \cos(3\theta+\theta_\xi) \nonumber\\
&&- \lambda_{\phi \xi 2} |\phi| v_\xi^3 \cos (\theta-3\theta_\xi)
 - \lambda_{\phi \xi 3} \mu |\phi|^2 v_\xi \cos (2\theta-\theta_\xi) \nonumber\\
&& - \lambda_{\phi \xi 4} \mu |\phi| v_\xi^2 \cos( \theta + 2\theta_\xi ) + \cdots
\end{eqnarray}
where the dots represent terms obtained after integrating out $h_\xi$, e.g., $\phi^5$, $(\phi^2 e^{-i\theta_\xi})^2$ and $(\phi^3 e^{i \theta_\xi})^2$. The final VEVs are given by $\langle \phi \rangle = v_0 e^{i 2\pi k/5}$ and $\langle \xi \rangle =v_\xi e^{i 4\pi k /5 }$ for $k=0$, 1, ..., 4, with $v_0$ the solution of $\partial V_{Z_5}^{\rm eff} / \partial \phi|_{\theta = \theta_\xi=0}=0$.

\subsection{Domain walls from $Z_6$ breaking}

In $Z_6$ symmetry, we focus on the two examples in C2) and C3). In C2), $\phi$ and $\xi$ respectively have charges $1$ and $3$ and the potential is given in Eq.~\eqref{eq:V_C2}:
\begin{eqnarray*}
V_{Z_6}^{\rm C2} &=& -\mu^2 \phi^* \phi + \lambda_1 (\phi^* \phi)^2 - \frac{1}{2}\mu_\xi^2 \xi^2 + \frac{1}{4}\lambda_\xi \xi^4 \\
&&- \lambda_{\phi \xi} (\phi^3+\phi^{*3}) \xi \,.
\end{eqnarray*}
Although the whole symmetry is $Z_6$, the charged-3 field $\xi$ transforms only in its subgroup $Z_2$ and hence can be treated as a real field. Once $\xi$ gains a VEV $\langle \frac{\xi}{\sqrt{2}} \rangle = \pm  v_\xi$ (with $v_\xi = \sqrt{\mu_\xi^2/2 \lambda_\xi}$), the sub $Z_2$ symmetry is broken, and a residual $Z_3$ is left. This step of SSB leads to the generation of $Z_2$ domain walls. $\xi$ in each cosmic cell wrapped by the wall takes a VEV at either $+v_\xi$ or $-v_\xi$. Inside a cell with the $+v_\xi$ VEV, the potential  is left in the $Z_3$-invariant form
\begin{eqnarray} \label{eq:V_C2_Z3}
V_{Z_3} &=& -\mu^2 \phi^* \phi + \lambda_1 (\phi^* \phi)^2  - \lambda_{\phi \xi} v_\xi (\phi^3+\phi^{*3}) \,.
\end{eqnarray}
As discussed before, $\phi$ has three degenerate vacua, $\langle \phi \rangle = \frac{\mu}{\sqrt{2\lambda_1}} (\beta + \sqrt{1+\beta^2}) \, e^{i2\pi k /3}$ for $k=0$, 1, 2. These vacua are separated by $Z_3$ domain walls formed following the SSB of $Z_3$. In the cell with the $-v_\xi$ VEV, the potential of $\phi$ is given by Eq~\eqref{eq:V_C2_Z3} with a sign difference of the last term. VEVs of $\phi$ are given by $\langle \phi \rangle = \frac{\mu}{\sqrt{2\lambda_1}} (\beta + \sqrt{1+\beta^2}) \, e^{-i2\pi k /3}$ for $k=0,1,2$, and they are separated by the $Z_3$ domain walls.

In summary, in the C2) case we have $Z_2$ domain walls from the $\xi$ field separating the Universe into $\langle\xi\rangle=+v_\xi$ and $-v_\xi$ cells, while in each cell the $Z_3$ domain walls from the $\phi$ field further separate the space into $\langle \phi \rangle = \frac{\mu}{\sqrt{2\lambda_1}} (\beta + \sqrt{1+\beta^2}) \, e^{\pm i2\pi k /3}$ cells. The whole picture of the Universe is like a {\it pomegranate} that it is split into a few chambers by membranes and each chamber is further divided into cells by another type of membranes.
Note that VEV of $\phi$ induces a shift of the two VEVs of $\xi$, but does not induce any bias between them. Therefore, the pomegranate-like defects keep stable at high scales before explicit breaking terms dominate the potential. 

In C3), $\phi$ and $\xi$ have charges $2$ and $3$. As seen from the potential in Eq.~\eqref{eq:V_C2}, there is no cross coupling between $\phi$ and $\xi$. The VEVs of  $\phi$ and $\xi$ leads to $Z_6$ spontaneously broken to $Z_2$ and $Z_3$, respectively. The SSB for $Z_6 \to Z_2$ generates $Z_3$ domain walls and that for $Z_6 \to Z_3$ generates $Z_2$ domain walls. Without considering explicit breaking, there will no interaction between $\phi$ and $\xi$ explicitly. These walls are transparent to each other. Their evolution should be independent from each other.

\subsection{Generalisation to $Z_N$ breaking}

We have found complicated domain walls when the SSB evolves multiple scalars in  a renormalisable potential.
Although we take $Z_5$ and $Z_6$ as examples, these complex domain walls exist in SSB of other Abelian discrete symmetries.
We recall categories C1), C2) and C3) in section~\ref{sec:more_scalars},
and generalise the discussions to $Z_N$ for $N \geqslant 5$.

In C1), all charges of scalars are coprime with $N$ in a $Z_N$ symmetry. Potential terms of each single scalar preserves $U(1)$ symmetry. The scalar which triggers the first-step of SSB breaks only the approximate $U(1)$ symmetry. Cross couplings between different scalars, which are the only sources breaking $U(1)$ explicitly, trigger the SSB of $Z_N$. The breaking chain is the same as in Eq.~\eqref{eq:chain1}, and we arrive at string-bounded adjacency walls after the two-step SSB.

In C2), we denote the greatest common divisor as ${\rm gcd} (q_\xi, N) = N_1$ (with $N_1> 1$ as required). $Z_N$  acting on $\xi$ behaves as a $Z_{N/N_1}$ acting on a scalar field with charge $q_\xi/N_1$.  $Z_{N/N_1}$ is broken after $\xi$ gains a VEV, followed by the generation of $Z_{N/N_1}$ domain walls. The residual $Z_{N_1}$ symmetry is conserved in each vacua wrapped by the walls. It is broken later after $\phi$ gains the VEV, leaving $Z_{N_1}$ walls wrapped by $Z_{N/N_1}$ walls.

In C3), all charges of scalars are coprime with each other, but have non-trivial common divisor with $N$. These charge assignments forbid terms $\xi^{n_\xi} \phi^{n_\phi}$ where $n_\xi$ and $n_\phi$ are positive integers. This property is simply proven as there is no positive-integer solution for the equation
\begin{eqnarray}
q_\xi n_\xi + q_\phi n_\phi = 0\, ({\rm mod} \,N)\,,
\end{eqnarray}
As no cross couplings between different scalars, $\xi$ and $\phi$ gain VEVs independently and break $Z_N$, to $Z_{{\rm gcd}(q_\xi, N)}$ and $Z_{{\rm gcd}(q_\phi, N)}$, respectively. The resulting $Z_{N/{\rm gcd}(q_\xi, N)}$ and $Z_{N/{\rm gcd}(q_\phi, N)}$ domain walls are  transparent with each other.

All these topological defects, together with those from the single scalar case, are summerised in Table.~\ref{tab:domain_walls}. In the multiscalar case, although we have discussed only two scalars, it is straightforward to extend the discussion by including more scalars into the picture.

\begin{table*}
\begin{center}
\begin{tabular}{| l | c | c | c | c |}
 \hline\hline
\multicolumn{2}{|c|}{Potential forms} & breaking chains & \multicolumn{2}{c|}{textures of domain walls} \\ \hline
\multirow{2}{*}{single scalar} & large $\phi^N$ & $Z_N \to 1$ & \cellcolor{yellow!50} adj. walls & \cellcolor{green!50} non-adj. walls ($N \geqslant 4$)  \\\cline{2-4}
& small $\phi^N$ & appr. $U(1) \to Z_N \to 1$ & \multicolumn{2}{c|}{\cellcolor{red!30} string-bounded adj. walls}   \\\hline
\multirow{3}{*}{multiscalar} & C1 & appr. $U(1) \to Z_N \to 1$ & \multicolumn{2}{c|}{\cellcolor{red!30} string-bounded adj. walls} \\\cline{2-4}
 & C2 & $Z_N \to Z_{{\rm gcd}(q_\xi, N)} \to 1$ & \multicolumn{2}{c|}{\cellcolor{blue!20}walls wrapped by walls} \\\cline{2-4}
 & C3 & $Z_N \to \Big\{\begin{array}{c}Z_{{\rm gcd}(q_\xi, N)} \\ Z_{{\rm gcd}(q_\phi, N)} \end{array}\Big.$ &  \multicolumn{2}{c|}{\cellcolor{purple!20}walls blind among diff. types}  \\
 \hline\hline
\end{tabular}
\end{center}
\caption{Incomplete classifications of domain walls generated from SSB of $Z_N$-invariant theories for given potential forms. In the multiscalar case, we have considered only two scalars $\xi$ and $\phi$, and $\xi$ gains the VEV early than $\phi$, triggering the first step of SSB.}
\label{tab:domain_walls}
\end{table*}

\section{Conclusion \label{sec:summary}}

$Z_N$ domain walls with $N\geqslant 3$ are predicted in new physics involving spontaneous symmetry breaking of $Z_N$ symmetries. The latter is critical in solving the strong CP problem, addressing the quark and lepton flavour puzzles and restricting interactions in supersymmetry. Understanding these domain walls provides us a complementary view to understand intrinsic physics behind. 
In this paper, we studied properties of $Z_N$ domain wall structures, in particular the $N=3,4,5,6$ cases. Main properties are summarised below.

In the $Z_3$ case, as all three degenerate vacua are adjacent to each other in the field space, only adjacency walls form. Taking the renormalisation into consideration, there is only a $\phi^3$ term breaking $U(1)$. The tension and thickness of the wall are $\sigma \sim m_a v_0^2$ and $\delta \sim m_a^{-1}$ where $v_0$ is the absolute value of the VEV and $m_a$ is the mass of the pseudo-Nambu-Goldstone boson after the SSB of the approximate $U(1)$. In the case of small $\phi^3$ terms,  we obtain topological defects of domain walls bounded by strings: cosmic strings form during the SSB of $U(1)$, and later during the SSB of $Z_3$, domain walls forms attaching to the string.

In the $Z_4$ case, all types of domain walls discussed in the $Z_3$ case could be generated from the SSB of $Z_4$. Numerical simulation shows that adjacency walls satisfy again the correlations $\sigma \sim m_a v_0^2$ and $\delta \sim m_a^{-1}$.  Beyond and what is more important, we observed and discussed properties of non-adjacency walls for the first time. These walls are defined via the domain wall solutions separating two vacua which are non adjacent in the field space, e.g., $v_0$ and $v_2$. We observed that a non-adjacency wall is stable only for large $\phi^4$ terms, which is the only renormalisable $U(1)$-breaking term in $Z_4$. In the case of small $\phi^4$ terms, a non-adjacency wall stores energy greater than twice of energy stored in an adjacency wall. Then the non-adjacency wall splits into two adjacency walls and a new vacuum is generated between the two adjacency walls. We further solve two marginal cases analytically in the appendix. One is the stability limit of vacua, and the other describes the marginal place where the non-adjacency wall is becoming unstable. In the latter case, we explicitly prove that the tension of a non-adjacency wall is twice of that of an adjacency walls.

To derive domain walls from larger $Z_N$ (for $N \geqslant 5$) symmetry breaking, one could consider the approach of either effective potentials with non-renormalisable $\phi^N$ terms or including one more scalar, e.g., $\xi$, with suitable charge alignements. The former approach preserves an approximate $U(1)$ symmetry and gives topological defects no more than domain walls bounded by strings. In the latter case, depending on $N$ and charges arranged for scalars, additional interesting topological defects may be generated as summarised below.

In the $Z_5$ case, we arrange charges of $\phi$ and $\xi$ to be $1$ and $2$ without loss of generality. Renormalisation condition forbids any $U(1)$-breaking terms in the potential of any single scalar and thus, terms breaking $U(1)$ but preserving $Z_5$ can be generated via cross couplings of $\phi$ and $\xi$. In this case, once these two scalars gain VEV at different energy scale, the SSB of a $U(1)$ symmetry always happens earlier than that of $Z_5$. We again arrive at domain walls bounded by strings. This case is generalised to $Z_N$ with scalar charges $q_\phi$ and $q_\xi$ which satisfy ${\rm gcd}(q_\xi, N) = {\rm gcd}(q_\phi, N) = 1$, where gcd is the abbreviation of the greatest common divisor.

In the $Z_6$ case, we found two more different structures of domain walls which may be generated from the SSB of $Z_6$ with respect to the charge alignment of the scalars. For charges assigned as $q_\phi = 1$ and $q_\xi = 3$ and $\xi$ gains VEV earlier than $\phi$, $Z_2$ domain walls generated first induced by the VEV of $\xi$. Each vacua of $\xi$ satisfies a residual $Z_3$ symmetry. After $\phi$ gains the VEV, $Z_3$ domain walls are generated, wrapped by $Z_2$ domain walls. This ``walls wrapped by walls'' structure can also be generated in other $Z_N$ symmetries if conditions ${\rm gcd}(q_\xi, N) > 1$ and ${\rm gcd}(q_\phi, N) = 1$ are satisfied.
On the other hand, if charges are arranged as $q_\phi = 2$ and $q_\xi = 3$ in $Z_6$, no cross couplings between scalars exist, $Z_3$ domain walls and $Z_2$ domain walls resulted from $\phi$ and $\xi$ are generated independently, and they are transparent with each other. This kind of walls can be generated in the general case that ${\rm gcd}(q_\phi, N),\, {\rm gcd}(q_\xi, N) > 1$ and ${\rm gcd}(q_\phi, q_\xi) = 1$ are satisfied.

In Table~\ref{tab:domain_walls}, we summarise all these Abelian domain walls as discussed in the paper. Due to the rich structures of domain walls, we expect they will have interesting cosmological applications, which will be carried out in our coming works.

\section*{Acknowledgement}

K.P.X. is supported by the National Science Foundation under grant number PHY-1820891, and PHY-2112680, and the University of Nebraska Foundation. Y.L.Z. acknowledges National Natural Science Foundation of China under grant No. 12205064.

\appendix

\section{Domain wall properties in special cases in $Z_4$}\label{app:Z4}

Domain walls from $Z_4$ breaking have a lot of interesting features which are not taken by $Z_2$ or $Z_3$ domain walls. In some marginal cases, the domain wall solution can be analytically solved. In this appendix, we will discuss the analytical solutions in two special cases. The first one is in the limit $\beta \to 1$, which is the limit of the vacuum stability. The second case is at $\beta = 1/3$, which describes the marginal place where non-adjacency walls are becoming unstable.

\subsection{In the limit $\beta \to 1$}
We discuss the domain wall solution in the adjacency branch with $\beta \to 1$. The EOMs of $h$ and $a$ are given in Eq.~\eqref{eq:EOM_Z4} and the BCs are given thereafter. As $1-\beta\ll 1$, the VEV takes a large value as $|v_0| = |v_1| \propto 1/\sqrt{1-\beta}$. The saddle point $|s_0| \approx 1/\sqrt{1+\beta}$ is much loser to the origin of the $\bar{h}$-$\bar{a}$ plane. One can imagine that the path, which begins from $v_0$, passes through the saddle point $s_0$, and arrives at $v_1$, follows almost a broken line with a right angle in the $\bar{h}$-$\bar{a}$ plane. With the help of this brief picture, we can split the path into two halves $0<-\bar{z}<+\infty$ and $0<\bar{z}<+\infty$. Profiles of $\bar{h}$ and $\bar{a}$ satisfy the permutation relation $\bar{h}(\bar{z}) = \bar{a}(-\bar{z})$. Once profiles of two scalars in one half path  are obtained, their profiles in the other half path are obtained following the permutation relation. Below, we will just focus on the solution in the second half path, i.e., $\bar{h}(\bar{z})$ and $\bar{a}(\bar{z})$ for $0<\bar{z}<+\infty$.

In the range $0<\bar{z}< \infty$, $\bar{h} < 1 \ll \bar{a}$ is satisfied. One can simply ignore the contribution of $\bar{h}$ to the wall, and consider only $\bar{a}$. The EOM of $\bar{a}$ approximately given by
\begin{eqnarray} \label{eq:EOM_Z4_1}
\bar{a}''(\bar{z}) &\approx& [-1+(1-\beta)\bar{a}^2] \bar{a} \,,
\end{eqnarray}
and the BCs are $\bar{a}(\bar{z}\to 0) \approx 0$ and $\bar{a}(\bar{z}\to +\infty) \approx 1/\sqrt{1-\beta}$. The solution satisfies the tanh function,
\begin{eqnarray}
\bar{a}(\bar{z}) \approx \frac{1}{\sqrt{1-\beta}} \tanh \big( \frac{\bar{z}}{\sqrt{2}} \big) \,.
\end{eqnarray}
Thus, the thickness and tension of the wall are obtained as
\begin{eqnarray}
&&\bar{\delta} \approx \sqrt{2} \,;\nonumber\\
&&\bar{\sigma} = 2 \int_0^\infty d\bar{z} \left[ \frac{1}{2} \bar{a}^{\prime 2} + \Delta \bar{V} \right] \approx \frac{2\sqrt{2}}{3(1-\beta)} \,.
\end{eqnarray}

\subsection{At $\beta =1/3$}

There is a critical value of $\beta$ that the non-adjacency wall begins to be unstable. We give an explicit proof that $\beta = 1/3$ is this value. Below we fix $\beta$ at this value. It is straightforward to check that each saddle point is collinear with two of non-adjacent vacua and is the middle point of the two vacua in the $\bar{h}-\bar{a}$ plane. In detail, $s_i=(v_i + v_{i+1})/2$ for $i=0$, 1, 2, 3, where $v_4 = v_0$ is identified.

We analytically calculate the domain wall solution with boundary $v_0$ and $v_1$ on the two sides. With the following parametrisation, $\bar{\varphi}_1=(\bar{h}+\bar{a})/\sqrt{2}$ and $\bar{\varphi}_2 = (\bar{a}-\bar{h})/\sqrt{2}$, Eq.~\eqref{eq:EOM_Z4} is decomposed to two decoupled equations
\begin{eqnarray} \label{eq:EOM_Z4_13}
\bar{\varphi}_m''(\bar{z}) =& \frac{4}{3} \left[ \bar{\varphi}_m^2 - (\frac{\sqrt{3}}{2})^2 \right] \bar{\varphi}_m\,,
\end{eqnarray}
for $m=1$, 2. The boundary conditions are rewritten to be $\bar{\varphi}_1|_{\bar{z}\to \mp\infty} = \sqrt{3}/2$ and $\bar{\varphi}_2|_{\bar{z}\to \mp\infty}= \mp \sqrt{3}/2$. The solution is simply given by
\begin{eqnarray}
\bar{\varphi}_1(\bar{z}) = \frac{\sqrt{3}}{2};\quad \bar{\varphi}_2(\bar{z}) = \frac{\sqrt{3}}{2}\tanh\left(\frac{\bar{z}}{\sqrt{2}}\right)\,.
\end{eqnarray}
The tension and thickness of the wall are obtained as $\bar{\sigma} = \sqrt{2}/2$ and $\bar{\delta} = \sqrt{2}$, respectively. Note that the tension here is half of $\sigma_{Z_2}(\beta=1/3) = \sqrt{2}$. Therefore, we have proven that the tension of a non-adjacency wall is twice of the tension of  an adjacency wall.

\bibliographystyle{apsrev4-2}

\end{document}